\documentclass[12pt,preprint]{aastex}

\newcommand{\eg}{e.g., }
\newcommand{\ie}{i.e., }
\newcommand{\cf}{cf.\ }

\begin{document}
\title{HST Imaging in the Chandra Deep Field South: II.
WFPC2 Observations of an X-Ray Flux-Limited Sample from the 1~Msec Chandra
Catalog%
\footnote{Based on observations made with the NASA/ESA Hubble Space Telescope,
which is operated by the Association of Universities for Research in Astronomy,
Inc., under NASA contract \hbox{NAS 5-26555}.}}

\author{Anton M. Koekemoer, Norman A. Grogin, Ethan J. Schreier}
\affil{Space Telescope Science Institute, 3700 San Martin Drive, Baltimore, MD 21218, USA}

\author{R. Giacconi, R. Gilli, L. Kewley, C. Norman, A. Zirm}
\affil{Department of Physics and Astronomy, Johns Hopkins University, Baltimore, MD 21218, USA}

\author{J. Bergeron, P. Rosati}
\affil{European Southern Observatory, Karl-Schwarzschild-Strasse 2, Garching, D-85748, Germany}

\author{G. Hasinger}
\affil{Astrophysikalisches Institut, An der Sternwarte 16, Potsdam 14482 Germany}

\author{P. Tozzi}
\affil {Osservatorio Astronomico di Trieste, Via G. B. Tiepolo 11, 34131 Trieste Italy}

\author{A. Marconi}
\affil{Osservatorio Astrofisico di Arcetri, Largo E. Fermi 5, 50125 Firenze, Italy}

\slugcomment{Accepted by {\it The Astrophysical Journal}, 15 October 2001}

\begin{abstract}
We present {\sl HST}/WFPC2 observations of a well-defined
sample of 40 X-ray sources with X-ray fluxes above the detection threshold of
the full 1~Msec Chandra Deep Field South (CDFS). The sensitivity and spatial
resolution of our {\sl HST} observations are sufficient to detect the optical
counterparts of 37 of the X-ray sources, yielding information on their
morphologies and environments. In this paper we extend the results obtained in
our previous study on the 300~ks CDFS X-ray data
	(Schreier et al. 2001,
Paper~I).
Specifically, we show that the optical counterparts to the X-ray sources are
divided into two distinct populations: 1)~an optically faint group with
relatively blue colors, similar to the faint blue field galaxy population,
and 2)~an optically brighter group, including resolved galaxies with average
colors significantly redder than the corresponding bright field galaxy
population. The brighter objects comprise a wide range of types, including
early and late type galaxies, starbursts, and AGN. By contrast, we show that
the faint blue X-ray population are most consistent with being predominantly
Type~2 AGN of low to moderate luminosity, located at higher redshifts
($z \sim 1-2$). This conclusion is supported by luminosity function models of
the various classes of objects. Hence, the combination of deep X-ray data with
the high spatial resolution of {\sl HST} are for the first time allowing us to
probe the faint end of the AGN luminosity function at cosmologically
interesting redshifts.
\end{abstract}

\keywords{X-rays: galaxies ---
	galaxies: active ---
	galaxies: evolution ---
	galaxies: high-redshift ---
	surveys}

\clearpage

\section{Introduction}

A principal question in astrophysics today involves understanding the formation
and evolution of galaxies. Central to this is the connection between large
scale structure formation, hierarchical mergers, and the presence of
supermassive black holes at the centers of most galaxies
	(\eg Magorrian et al.~1998).
Addressing these issues requires detailed multi-waveband observations of a
variety of galaxy classes, including normal galaxies, starbursts and active
galactic nuclei (AGN). Deep X-ray surveys in recent years have
provided a wealth of new information on the properties of AGN and starbursts,
and their role in galaxy evolution
	(\eg Hasinger et al.~1998,
			1999;
	Miyaji et al.~2000).
For example, the observed strong evolution of the low end of the AGN luminosity
function at high redshifts is likely related to an increased rate of
interactions and distorted/irregular morphologies in the early universe
	(Lilly et al.~1998;
	Abraham et al.~1999).
Recent ultra-deep X-ray surveys with Chandra have extended such results to
fainter fluxes and harder X-ray energies
	(cf.~Giacconi et al.~2001a,
			2001b;
	Brandt et al.~2001;
	Hornschemeier et al.~2001;
	Mushotzky et  al.~2000;
	Tozzi et al.~2001).
Further evidence of a connection between AGN and galaxy evolution has been
provided by the strong correlation between black hole mass and host galaxy
bulge velocity dispersion
	(Gebhardt et al.~2000;
	Ferrarese \& Merritt 2000).
However, the fundamental question of whether nuclear activity drives galaxy
evolution, or vice versa, still remains open, and can only be addressed by
obtaining high-resolution multi-wavelength observations of AGN {\it and} their
hosts over a wide range of redshifts. The combination of {\sl HST} and Chandra
provides a unique opportunity to address these issues by first detecting
distant AGN at X-ray energies, and subsequently obtaining high-resolution
{\sl HST} images of their host galaxy morphologies and environments.

In our first paper in this series
	(Schreier et al 2001, hereafter Paper~I),
we presented preliminary results from our program using the {\sl HST} Wide Field
Planetary Camera~2 (WFPC2) to image three regions in the Chandra Deep Field
South (hereafter CDFS;
	Giacconi et al.~2001a).
The analysis in that paper was based on the X-ray catalog and results from the
first 300~ksec Chandra exposure
	(Tozzi et al.~2001).

In the present paper, we extend our discussion to the full 1~Msec X-ray dataset
	(Giacconi et al ~2001b).
This provides a well-defined X-ray flux limited sample of objects that have
optical counterparts observed with {\sl HST}. We also present a more detailed
discussion of our {\sl HST}/WFPC2 observing strategy and analysis.
In Section~2 we describe the observational strategy used with {\sl HST}, in
Section~3 we discuss the techniques used to reduce the data, in Section 4 we
present the results from the observations, and in Section~5 we present a
discussion of the implications of the various optical properties of the X-ray
sources. Throughout this paper we adopt a cosmology with
$\Omega_{\rm M} = 0.3$, $\Omega_\Lambda = 0.7$, and
$H_0 = 70$~km~s$^{-1}$~Mpc$^{-1}$.

\section{Observing Strategy}

The CDFS is centered at R.A. 03 32 28, Dec. $-$27 48 30 (J2000), with deep
X-ray data covering an area $\sim\,0.1$~deg$^2$
	(Giacconi et al. 2001a,
			2001b).
We used {\sl HST}/WFPC2 to observe three regions within the CDFS, during
$22 - 27$~July 2000. Each of the three regions was observed for a total of five
orbits, comprising $\sim\,2$~orbits in the F606W filter and $\sim\,3$~orbits in
F814W. These two filters provide an optimal combination of high throughput,
wide bandwidth and minimal spectral overlap, with well characterized
photometric accuracy; the spectral coverage is continuous between
$\sim 5000 - 9500$~\AA, with a small overlap between the filters at
$\sim 7200$~\AA. Thereby these observations were specifically designed to yield
good sensitivity to the color gradients and other morphological features that we
expected to observe in the moderate-redshift galaxies in these fields (\cf
	Williams et al. 1996;
	Casertano et al. 2000).
We chose a somewhat shorter exposure time for the F606W filter due to its
higher overall throughput, thereby aiming to match our sensitivity to isophotal
features on comparable spatial scales in both bands. In
Table~\ref{tab:obs-summary} we present a summary of the field coordinates
(specified as the location of the WFALL-FIX aperture reference position on
WFPC2, which we used to obtain the observations), as well as the total exposure
times and position angle of the +V3 axis of the telescope.

Several instrumental issues need to be addressed when observing with
{\sl HST}/WFPC2; these include the removal of cosmic rays, the presence of
hot pixels and other detector artifacts, and the fact that the {\sl HST} point
spread function (PSF) is substantially undersampled by the
$\sim\,0\farcs1$~pixels in the three WF cameras (WF2, WF3 and WF4). We
therefore obtained the observations in each filter using the ``dither'' mode,
which involves offsetting the telescope by small increments to move the sources
around the detector. We restricted the offsets to a maximum of
$\lesssim\,0\farcs75$ along each axis (corresponding to a total of 7.5 WF
pixels). The use of larger shifts would have improved the cosmetic appearance
of the final combined dataset, particularly by increasing the counts in the
inter-chip gaps between the cameras. However, the $\sim 2\%$ geometric
distortion present across each camera introduces increasingly significant
non-uniform subsampling across the chip for offsets much larger than
$\sim\,10$~pixels. Since our primary objective was to obtain consistent
sub-sampling across each chip, we therefore used the minimal offsets.

We employed a 4-point parallelogram dither pattern, in which the telescope was
moved in steps of integral plus 1/2-pixel increments along both axes of the
detector. The integral pixel shifts mitigate the effect of bad pixels, while
the additional 1/2-pixel increments allow subsampling of the PSF. The ratio of
pixel scales between the PC and the three WF cameras is $\sim\,0.45$, thus an
offset of $0\farcs25$ corresponds to a 2.5-pixel shift in each of the three WF
cameras, and a 5.5-pixel shift in the PC. The 4-point parallelogram dither
pattern is therefore designed so that sources are optimally subsampled by all 4
quadrants of a pixel. We changed filters at each dither location, to minimize
the number of offsets that the telescope was required to perform and also to
reduce overhead. At each of the four dither locations we further obtained two
exposures to improve removal of cosmic rays, as well as providing some
redundancy to ensure that subsampling was achieved with at least one exposure
in each of the 4 pixel quadrants, for most pixels across each chip. The
multiple dithered exposures also reduce photometric errors due to flatfielding
uncertainties.

In Table~\ref{tab:dither} we show the fractional pixel shifts in all 4 cameras,
for each of the offsets. The dither offsets were specified in the POS TARG
reference frame of the WFALL-FIX aperture of WFPC2, and therefore indicate the
direction of motion of the field along the X and Y axes of the WF3 detector.
The orientation of the shifts for the other cameras rotates, corresponding to
the 90 degree rotational differences between the reference frames of the
various chips.

In Figure~\ref{fig:dither} we show the offsets that were achieved for each of
the three fields. Since each field was observed for five contiguous orbits, a
full guide star acquisition was only required at the start of each set of five
orbits, while a re-acquisition was carried out at the start of each of the
subsequent four orbits. For each of the three fields the observations were
obtained in FINE LOCK mode, with successful acquisition of both guide stars.
This permitted accurate offsetting and tracking during each set of five orbits.
We measured an overall offsetting precision of $\sim\,10$~mas r.m.s. when
comparing the actual offsets with those that were commanded to the telescope.
This translates to a fractional pixel positioning accuracy of $\sim\,0.1$~pixel
r.m.s. on each of the three WF chips, sufficient for our desired
$\sim\,1/2$-pixel subsampling. The corresponding fractional pixel offset r.m.s.
is larger on the PC camera due to its smaller pixel size, but this is mitigated
by the fact that the PC pixels provide much better sampling of the PSF.
Figure~\ref{fig:dither} confirms that the WF fractional pixel offsets
generally lie within their respective pixel quadrants.

In Table~\ref{tab:obs-detailed} we present a summary of the observations,
showing how the exposures in each filter were divided in terms of their dither
pattern location. Some asymmetries in the exposure times at the different
dither locations had to be introduced as a result of dividing four dither
positions into five orbits, along with our requirement for obtaining a longer
overall exposure time in F814W than F606W. However, we aimed to maintain
exposure times that were as uniform as possible at each dither location. Thus
the F814W exposure times varied between $600 - 800$~s with a mean of 725~s, and
the F606W exposures varied between $400 - 700$~s with a mean of 463~s. This
variation does not significantly compromise the statistical noise properties of
the final images.

\section{Data Reduction and Analysis Procedures}

\subsection{Pipeline Processing}

The observations were initially processed with the standard pipeline
calibration, at the point when they were retrieved from the HST Data Archive.
This procedure consists of the following steps: masking bad pixels;
analog-to-digital conversion; subtraction of the mean bias level; subtraction
of the ``superbias'' image; subtraction of the ``superdark'' image (scaled
according to the darktime of the observation); and multiplying by the flat
field reference image for the appropriate filter. These standard procedures
are documented in further detail in the HST Data Handbook V3
	(Voit et al. 1997).
After the data had been retrieved, they were recalibrated a few weeks later
when the most up-to-date dark calibration reference files for the time of the
observation became available.

\subsection{Dithered Image Registration}

After calibration, the images were processed using tasks in the IRAF/STSDAS
{\sl dither} package. The steps outlined in this section are similar to those
used recently for the {\sl HST}/WFPC2 observations of the HDF-S
	(Casertano et al. 2000),
and also described in substantially more detail in the HST Dither Handbook
	(Koekemoer et al. 2000).
Here we present a summary of the steps involved, and further details can be
obtained by consulting these references.

The first step in this procedure involved removal of the sky background from
each of the separate chips, using the task {\sl sky} to bin the pixel fluxes
into a histogram and calculate the mean background based on the width of the
histogram distribution about its peak value. After sky subtraction, new pixel
flux histograms were constructed to verify that the peak of the histogram
distribution was now centered at zero for each chip.

We chose to use cross-correlation to calculate the relative offsets between
images, since the F606W and F814W filter images of all three fields generally
contained a sufficient number of objects bright enough to enable successful
cross-correlation. The results from the cross-correlation were later verified
by examining the jitter files for each observation retrieved from the HST Data
Archive, and this generally revealed good overall agreement between the two
methods (better than 10~mas).

To prepare the images for cross-correlation, the task {\sl precor} was run on
each chip. This task essentially filters the image, producing an output image
that has flux in pixels only for sources having a given number of adjacent
pixels above some specified value. This strategy allowed effective removal of
most cosmic rays, by setting the required number of adjacent pixels
sufficiently high. In the output image, blank regions and very faint objects
were set to zero, in order to avoid adding noise to the cross-correlation.

Next, the task {\sl offsets} was run, which makes use of the program
{\sl crosscor} to perform the cross-correlation. Each dither location
contained 8 exposures for each of the two filters. The first of the 8 images
was defined as the reference image, and {\sl crosscor} then produced
cross-correlation images for the other pointings relative to this image.

The task {\sl shiftfind} was then run on the output cross-correlation images
from {\sl crosscor}, fitting a two-dimensional Gaussian to the central
cross-correlation peak and yielding as output the x and y pixel location of the
peak, together with their 1-$\sigma$ uncertainties. Finally, the task
{\sl avshift} was run on the shifts of all 4 WFPC2 chips, using known
information about the geometric relationships between the chips to average the
shifts that were obtained independently for each chip, thereby producing a
final improved value of the x,y shift of each WFPC2 pointing in the series
of 8 images (these are the shifts displayed in Figure~\ref{fig:dither}).

\subsection{Cosmic Ray Rejection and Image Combination}

After determining the shifts, the next steps involved combining the images and
simultaneously carrying out cosmic ray rejection. To do this, we made use of a
new capability of the {\sl drizzle} software
	(Fruchter \& Hook 2001),
which permits direct removal of cosmic rays from individual images while at the
same time performing the steps involved in combining them. We adopted an output
image scale of 0.5~input WF pixels, motivated by the general success of our
observational strategy in achieving half-pixel subsampling across all three WF
cameras. While we obtained two images at each of the 4 dither positions, this
was primarily intended to ensure redundancy in the subsampling; any attempt to
remove cosmic rays by simply combining the two images would not result in 
optimal cosmic ray removal. Therefore we adopted a strategy of treating all 8
images separately in the cosmic ray removal process outlined here.

To create the cosmic ray masks, we first ran {\sl drizzle} on each input image,
using the offsets determined previously to shift them all onto separate output
images having a common alignment. These images were then combined using a robust
median rejection technique to remove all the cosmic rays, producing a somewhat
noisy but clean image containing only real astronomical sources. This image was
then transformed back onto the original frame of each input image, using the
task {\sl blot}, which essentially reverses the shifts that were applied by
{\sl drizzle}. The tasks {\sl deriv} and {\sl driz\_cr} were then run on these
images, thereby creating a final mask for each input image that contained all
the cosmic rays, bad pixels, and other defects.

After creating the masks, the final step involved running {\sl drizzle} once
more on the input images, using the same shifts, but with the difference being
that all the input images were ``drizzled'' onto a single output image, and
also making use of the mask images to prevent cosmic rays and other defects
from propagating onto the output image. The {\sl drizzle} program also creates a
corresponding ``weight'' image, representing the relative contribution of input
pixels to each output pixel. We shrunk the footprint of the input pixels by a
factor of 0.7 (specified by the {\sl drizzle} parameter {\tt pixfrac}), thereby
improving the resolution of the output image by reducing the overall scale with
which the output pixels are convolved, while still retaining sufficiently
uniform sampling of the input image plane by the output pixel grid. More
detailed discussions of these effects are presented in
	Koekemoer et al. (2000)
and
	Fruchter \& Hook (2001).
We also created another set of images with {\tt pixfrac} set to 1, which yielded
somewhat lower resolution but improved weighted image statistics for object
detection and photometry. These images were used to carry out the source
detection described in the next section, while the higher resolution images
with {\tt pixfrac} set to 0.7 were subsequently used to examine the detailed
morphology of the detected X-ray sources, as presented in this paper.

\subsection{Optical Identification of the X-ray Sources\label{sec:opt_id}}

We first created a detection image for each WFPC2 field by adding the F606W and
F814W images, and then extracted objects using the SExtractor program. The
measured count-rates were converted to the {\sl HST}/WFPC2 ABMAG photometric
system using the best available information
	(cf.~{\sl HST}/WFPC2 Instrument Handbook, V.5, 2000).
Our source detection and extraction procedures are described in Paper~I. In a
forthcoming paper (Grogin et al. 2001, in preparation) we will describe in
more detail our photometric procedures, including radial surface brightness
photometry and profile fitting.

In Figure~\ref{fig:fullfields} we present greyscale images of the three fields,
made by combining the F606W and F814W datasets. We also overlay the locations
of all the X-ray sources in the 1~Msec catalog that are within the field of
view of our WFPC2 images. Our three WFPC2 fields contain a total of 40 X-ray
sources listed in the 1~Msec Chandra catalog, with $12-15$ X-ray sources per
field.

We registered each of the three WFPC2 fields independently to the {\sl CXO}
reference frame by subtracting the error-weighted median offset of the nearest
{\sl HST}/WFPC2 optical counterpart to each {\sl CXO} source. The overall
offset is different for each of the three fields since different guide stars
were used --- the general size of the offsets ($\sim\,1$\arcsec) is consistent
with the uncertainties inherent in the {\sl HST} guide star catalog, as well as
additional astrometric uncertainties present in the Chandra image.

After registering the frames by removing the overall offsets, we next examined
the positional offset between each {\sl CXO} source position and the location
of the nearest {\sl HST} source. The mean offset between the X-ray and nearest
optical source was found to be $0\farcs33$ in right ascension and $0\farcs38$
in declination. When we examined the distribution of positional offsets in
right ascension and declination combined, we found a 1-$\sigma$ r.m.s.
deviation of $0\farcs5$, which is consistent with the general {\sl CXO}
positional uncertainties quoted by
	Tozzi et al.~(2001),
particularly given that the {\sl CXO} PSF becomes substantially degraded at the
locations of some of our WFPC2 fields, away from the CDFS field center. 
In Figure~\ref{fig:offsets}(a) we show the positional offsets between the X-ray
sources and the nearest detected optical counterparts, for each X-ray source,
and in Figure~\ref{fig:offsets}(b) we show the magnitude of the offset as a
function of off-axis angle in the Chandra ACIS-I image. The scatter in the
offsets remains approximately constant and does not show any strong increase at
larger off-axis angles. We found that 37 of the 40 X-ray catalog sources
have an optical counterpart detected within our 3$\sigma$ positional tolerance
from the X-ray location. We note that in some cases, more than
one optical identification is located within $0\farcs5$ of the X-ray source,
hence there remains a formal statistical uncertainty in these cases about which
optical source is the true counterpart. However, in these cases, a comparison
between the X-ray and optical image generally reveals that the X-ray contours
are well centered on a single object, as we discuss further in
Section~\ref{results} where we present the X-ray overlays on the optical
images.

The remaining three X-ray sources are more than 2\arcsec\ away (\ie
$\gtrsim 4$$\sigma$) from the nearest {\sl HST}/WFPC2 object above our
detection threshold. All three of these sources are on the PC and have
$3\sigma$ detection limits of F606W$\,\geq\,$26.9 and F814W$\,\geq\,$26.4,
based on 0\farcs5-radius apertures centered on the {\sl CXO} coordinates. We
note, however, that the detection limits on the PC are shallower than on the WF
chips, primarily resulting from the increased read-noise due to the fact that
there is a higher number of the smaller PC pixels per unit area on the sky.
When we examine the fluxes of our remaining sample of 37 objects, we find that
35 of these (94.6\%) would still have been detectable by the PC chip. Thus, for
our total sample of 40 sources, the detection efficiency of the PC would be
87.5\%, which is not statistically different from the 94.6\% efficiency
previously inferred (given the sample size of 40 sources). Since the PC field
of view covers only $\sim 6\%$ of our total area, we consider it somewhat
unlikely that the three non-detections on the PC could form part of a
completely different population that is much fainter than any of the other
sources that are detected on the WF chips.

To explore this issue further, however, we also examine the X-ray properties of
the three undetected sources, as compared with the rest of the sources. Their
average flux is $2.14 \times 10^{-16}$~ergs~s$^{-1}$ and
$1.32 \times 10^{-15}$~ergs~s$^{-1}$ in the soft and hard band respectively. Of
the optically identified sources, 19 have total X-ray fluxes greater than this,
while 18 have fainter X-ray fluxes, thus the average flux of the three
optically undetected sources lies near the median of the X-ray flux
distribution. Of the 18 optical identifications that have fainter X-ray flux,
2 would not have been detected had they landed on the PC chip. Moreover, when
we examine the constraints on the X-ray/optical flux ratios of the three
undetected sources, we find that the lower limits on this ratio are not
inconsistent with the typical values for the remainder of the sources, as can
be seen by examining the X-ray fluxes and optical magnitudes in
Table~\ref{tab:all_sources}. Thus, although we cannot completely rule out the
possibility that these three objects may have a much higher X-ray/optical ratio
than the other sources in our sample, we consider it more plausible that they
would likely have been detected had they fallen on the WF chips instead of the
PC, and that these sources may thus form part of the same population as some of
the optically faint sources that are detected on the WF chips.

\section{Results\label{results}}

\subsection{Optical Properties of the X-Ray Sources}

Figure~\ref{fig:smallfields_grey} shows $20\arcsec\times20\arcsec$ {\sl HST}
subimages (derived from the combined F606W and F814W datasets), centered on the
coordinates of each of the 40 {\sl CXO} sources from the 1~Msec catalog
	(Giacconi et al. 2001b)
that are located within our WFPC2 fields, with contours overlaid from the 1~Msec
$0.5-7$~keV X-ray image.
It can be seen that, in general, the agreement between the centroid of the
{\sl CXO} contours and the peak of the optical emission is remarkably good.
This is the case not only for optically bright, unresolved objects but also
for many of the fainter objects as well as those that are extended in the
optical images but do not display a dominant nuclear component. We also note
that the X-ray contours in general are unresolved or display only marginal
spatial extent; the variation in the structure of the X-ray contours is
predominantly due to changes in the {\sl CXO} PSF across the fields of view
covered by our {\sl HST} observations
	(Giacconi et al.~2001b).

In Table~\ref{tab:all_sources}, we list the F606W and F814W magnitudes (or
3$\sigma$ upper limits) for all 40 {\sl CXO} sources located in our WFPC2
fields, as measured from our SExtractor photometry. We also tabulate the
{\sl CXO} fluxes in the soft ($0.5-2$~keV) and hard ($2-10$~keV) bands (columns
$F_{\rm XS}$ and $F_{\rm XH}$ respectively), together with the X-ray hardness
ratio values $HR$ reported by
	Giacconi et al.~(2001b)
(ranging from $HR=-1$ for sources detected only in the soft band, to $HR=+1$
for sources detected only in the hard band).

In Table~\ref{tab:all_sources} we also present information about the optical
morphology of the {\sl HST} counterparts, indicating whether or not the source
is resolved and if so, what galaxy class it appears to be. We found that the
SExtractor star/galaxy separation was reliable to F606W$\,\sim\,25.5$,
substantially fainter than the majority of the counterparts to the {\sl CXO}
sources. Thirty of the 37 optical counterparts to the X-ray sources are clearly
resolved, of which 25 were sufficiently extended to permit detailed examination
and classification as elliptical, spiral or irregular galaxies. The remaining
five were too faint and/or not sufficiently extended to permit reliable
classification, thus we list them as ``indeterminate'' morphology in
Table~\ref{tab:all_sources}.

\subsection{Color-Magnitude Properties of the X-ray Optical Counterparts}

In Figure~\ref{fig:col_mag} we plot the F814W magnitudes and F606W$\,-\,$F814W
colors of all the sources in our SExtractor catalog
	(Grogin et al. 2001, in preparation).
The field galaxies (unassociated with any X-ray source) are
plotted as small dots; the 37 optical counterparts of the {\sl CXO} sources
are plotted with larger symbols indicating their optical morphology and X-ray
properties. Circles represent resolved early-type galaxies (types S0 and
earlier); squares are resolved late-type spirals~/ irregulars (types Sa and
later); stars represent clearly unresolved sources; and triangles are objects
of indeterminate morphology. The sizes of the symbols correspond to total X-ray
flux in the $0.5 - 10$~keV band, with size being proportional to higher flux.
The symbols are also shaded with a greyscale intensity according to their
hardness ratio: soft sources are shown with a light shading, while hard sources
are darker.

With our increased number statistics, we are able to confirm the dichotomy
(reported in
	Paper~I)
in the color magnitude distribution of the optical counterparts of the X-ray
sources: a brighter group with F814W $\sim 18-24$, and a fainter group with
F814W$\,\gtrsim 24$. The fainter group is consistent in its color distribution
with the field galaxy population. The other group is both significantly
brighter and significantly redder, on average, than the population of field
sources.

We overlay predicted templates for elliptical, spiral, and irregular~/
starburst galaxies, as well as Type~1 AGN (as in Paper~I; see also
	Williams et al.~1996).
These tracks show that the population of bright, extended X-ray sources should
consist of a mix of morphological types at low to moderate redshifts, including
Type~1 AGN, while the optically fainter X-ray sources are likely too blue to be
early-type galaxies.

In order to further explore this apparent dichotomy in color-magnitude space,
we present in Figure~\ref{fig:smallfields_rgb} a montage of all the WFPC2
images of the X-ray sources, with blue representing F606W, red representing
F814W, and green a mean image of the two. The overall structure of this figure
is schematically similar to Figure~\ref{fig:col_mag}: the vertical panels are
arranged in order of decreasing apparent magnitude toward the bottom, with
redder objects toward the right; the horizontal panels at the bottom represent
the faint group (most of which are blue) together with the three X-ray sources
that have no optical counterparts in the {\sl HST} images. The sources are
divided into the following groups, according to their distribution in the
color-magnitude diagram presented in Figure~\ref{fig:col_mag}:
\begin{itemize}
\item Bright (F814W $\lesssim 24$); blue (F606W$\,-\,$F814W $\lesssim 0.7$):
These sources all contain an unresolved nucleus, contributing a substantial
portion of the total flux of the object -- they are likely Type~1 AGN, and this
is confirmed by the optical spectra as well as the X-ray and optical
luminosities of those objects for which spectra are available so far. More
specifically, we have so far obtained spectra for 4 of these objects (CDFS ID's
39, 60, 62, and 63), and all of them can be definitively classified as Type~1
AGN on the basis of their spectra. More detailed discussions concerning the
analysis and interpretation of the spectroscopic data will be presented in a
forthcoming paper
	(Hasinger et al. 2001, in preparation).
\item Bright (F814W $\lesssim 24$); intermediate color
(F606W$\,-\,$F814W $\sim 0.7 - 1.2$):
These objects are all spatially resolved by our {\sl HST} observations, and
we are able to classify them morphologically as either elliptical, S0, or
spiral galaxies. The fluxes of these objects are dominated by their host
galaxies rather than by a bright, unresolved nucleus; a faint nucleus may
however be present in a number of these objects. We further divide these
objects into soft (left-hand column) and hard (right-hand column); there
appears to be no discernible systematic difference between these classes in
terms of either the host galaxy morphological type, the prominence of the
central nucleus (or lack thereof), or apparent magnitude. However, these
sources are in general harder than the bright blue population.
\item Bright (F814W $\lesssim 24$); red (F606W$\,-\,$F814W $\gtrsim 1.2$):
The objects in this class are also spatially resolved in our data, and appear
to consist predominantly of early-type galaxies. Two of them display some hint
of spiral structure, but their overall color is still much redder than the
more spatially extended spirals in the bluer populations; this may be due
either to a higher redshift or intrinsically large amounts of dust. We again
divide these objects into soft (left-hand column) and hard (right-hand column),
and once more there is no clear morphological difference between the two types.
\item Faint (F814W $\gtrsim 24$):
These objects are plotted as horizontal panels near the bottom of the
figure, divided into hard (top panel) and soft (bottom panel). Although these
objects are small, they are all nonetheless resolved by our {\sl HST}
observations --- there is no indication of any unresolved nuclear component
contributing a substantial part of the flux. Thus we are able to rule out
the possibility that these are distant Type~1 AGN; furthermore Type~1 AGN
at these optical flux levels would have been too faint to be detected in the
current Chandra dataset. Instead, these objects appear to be drawn from the
same population as the rest of the faint blue galaxy population. We draw
attention to one exceptional object, which is much redder than the others and
presumably belongs to a different class; we will return to a more detailed
discussion of this source in the next section.
\end{itemize}
For completeness, we also show the three non-detections at the bottom of
Figure~\ref{fig:smallfields_rgb}). These were all on the PC chip where the
detection threshold is shallower than on the WF chips, as we have discussed
earlier in Section~\ref{sec:opt_id}.

\section{Discussion: Optically Bright and Faint Sources -- Two Distinct
	Populations}

The single most prominent result from this study is the clear separation of the
X-ray sources into two apparently distinct populations in color-magnitude space.
Our earlier study, based on the 300~ks data
	(Paper~I)
suggested this but contained only 5~sources in the faint population. By
extending our study to the full sample of X-ray sources from the 1~Msec
catalog, this result has now been shown to remain robust, with 10 faint sources
and 27 in the brighter group.

To quantitatively examine the nature of these two populations, we used
published SEDs and luminosity functions for various classes of galaxies to
calculate the expected numbers of objects detected in our study (as described
in
	Paper~I,
and also to be addressed in more detail in a forthcoming paper,
	Koekemoer et al. 2001, in preparation).
In Figure~\ref{fig:popns}, we present a histogram of the F814W magnitude
distribution of the X-ray sources, together with the predicted distributions
for ellipticals, spirals, starbursts, and AGN in the area of sky corresponding
to our three WFPC2 fields, and applying the appropriate X-ray detection
thresholds. We have not adjusted any free parameters in the models when
calculating the distributions, which could of course improve the normalizations
somewhat --- rather, we have simply adopted the published luminosity functions
as representative of each of the galaxy classes that we detect. Specifically,
the distribution of each galaxy class is simply the result of calculating the
number counts of that type of object based on the assumed SED, luminosity
function and evolution (which produces a distribution that rises steeply
toward fainter magnitudes), and applying the X-ray flux limits corresponding to
the 1~Msec survey. This begins to cut off the rising distribution at the
corresponding F814W magnitude, which to first order is characteristic of the
$F_X/F_{\rm opt}$ ratio of each class of object; second-order effects such
as redshift K-correction produce a tail toward fainter magnitudes. It is thus
encouraging that the models agree with the observed distributions to this level
of detail.

The models in Figure~\ref{fig:popns} suggest that the majority of the X-ray
optical counterparts are expected to be relatively bright, comprising a
wide range of galaxy types including ellipticals, spirals, starbursts, and
AGN, and this is completely consistent with what we observe.  We note that
many of the objects classified as ``spirals'' (from the SEDs published by
	Schmitt et al. 1997)
may in fact harbor low-luminosity, X-ray emitting AGN, as suggested by
	Ho, Filippenko \& Sargent (1995);
their X-ray luminosities are likely to be at least one to two orders of
magnitude below that of the objects classified by
	Schmitt et al. (1997)
as ``Seyferts'. Much more interesting, however, is the realization that the
population of fainter, blue X-ray emitting sources apparently consist
predominantly of Type~2 AGN located at higher redshifts ($z \sim 1 - 2$). This
interpretation is supported independently from a study recently completed by
	Alexander et al. (2001)
for the CDF-N, and here we further explore some of its implications.

The lack of Type~1 AGN in the population of optically faint X-ray sources is
primarily a consequence of the $F_X / F_{\rm opt}$ ratio, and can be understood
as follows. If we consider Type~1 and Type~2 AGN of comparable X-ray luminosity,
located in similar host galaxies at redshifts $z \gtrsim 1$, then the detected
X-ray emission from both types of objects would be comparable --- although some
of the softer X-rays from the Type~2 AGN would be absorbed by the obscuring
torus, the majority of the X-rays harder than a few keV are able to penetrate,
and at higher redshifts these X-rays are shifted toward the soft end, thereby
producing comparable total X-ray fluxes for Type~1 and Type~2 AGN in the
$0.5 - 10$~keV Chandra bandpass. Thus, we can consider the hard X-ray emission
to be approximately isotropic. However, the total optical emission from
a source containing a Type~1 nucleus is brighter than that of a comparable
Type~2 by a few magnitudes, because of the dominant contribution from
the unobscured Type~1 nucleus. Hence, Type~1 AGN detected near the X-ray
sensitivity threshold of the 1~Msec survey cannot be optically fainter than
F814W$\,\sim 22 - 24$. However, Type~2 AGN detected near the same X-ray
threshold can be considerably fainter in the optical since the nuclear emission
is obscured, and only the host galaxy contributes to the observed optical
emission. If the Type~2 AGN is of only moderate luminosity, as expected on the
basis of the X-ray fluxes presented here, then it is likely located in a
late-type galaxy, which at $z \sim 1 - 2$ would become optically
indistinguishable from the faint blue galaxy population. We note that more
distant, or less luminous, Type~1 AGN may of course still be present among the
faint blue galaxy population (F814W$\,\gtrsim 24$). They would be too faint at
X-ray energies to be detected at the depth of the 1~Msec survey, but in this
scenario they should become visible in deeper Chandra observations, for example
as currently planned for the CDF-N.

We can similarly rule out starbursts as a possible explanation for the faint
blue X-ray population, once again based on the ratio of X-ray to optical
emission which is substantially lower for starbursts than for Type~2 AGN. Given
the X-ray sensitivity thresholds of the 1~Msec catalog, starbursts detected in
this survey are not likely to be optically fainter than F814W$\,\sim 23$ at
most, since any fainter (more distant) starbursts would not be detectable in
the X-rays. The same arguments apply to ellipticals and spirals without AGN ---
those that are at sufficiently high redshift to be in the part of the diagram
with F814W$\,\gtrsim 24$ would not be detected in the 1~Msec survey. These
arguments are supported quantitatively by the comparison between the models and
the observed number distributions of objects in Figure~\ref{fig:popns}, which
imply that the predominant population of faint blue X-ray sources has to be low
to moderate luminosity Type~2 AGN, located at redshifts $z \sim 1 - 2$.

We finally describe one object of particular interest in the faint
population, CDFS ID \#515 (J033232.2$-$274652), located on the WF4 chip in
the CDFS1 field. This object is at least one magnitude redder than the rest of
the faint blue population; it is undetected in F606W down to a 3$\sigma$ limit
of 27.9, thus its color is F606W$\,-\,$F814W $\gtrsim 1.4$. If this object is a
Type~2 AGN then its host galaxy is early-type or excessively dust-rich.
However, we cannot rule out the possibility that this object could be an highly
reddened QSO --- \ie an AGN of much higher intrinsic luminosity than the other
active nuclei discussed in this section, but reddened by substantial amounts of
dust. The fact that we only observe one of these in our sample could support
the interpretation of a higher intrinsic luminosity, thus a lower space density,
for this object, although other explanations may also be possible. It is
expected that deep spectroscopic observations would yield more definitive
information about this object.

\section{Conclusions}

We have presented {\sl HST}/WFPC2 observations of an X-ray flux limited sample
of 40 sources from the 1~Msec CDFS catalog, thereby completing our earlier
study which had been based on the 300~ks X-ray catalog. We find that 37 of the
X-ray sources in our WFPC2 fields have apparent optical counterparts within
$\sim\,1$~arcsec; the three non-detections were all on the PC chip where the
detection threshold is shallower than on the WF chips, but their non-detection
is not inconsistent with the optical/Xray fluxes of the faint sources detected
on the WF chips at comparable magnitudes. In fact, had they fallen on the WF
chips we find that they would most likely have been detected; since all 37
sources on the WF chips are detected, we conclude that {\sl HST}/WFPC2
observations to this depth are potentially capable of detecting all the sources
from the 1~Msec catalog.

The principal result from this work, which had been suggested with a small
number of sources in Paper~I and now confirmed with a substantially increased
sample in the present paper, is the realization that the optical counterparts
of the 1~Msec X-ray sources consist of two distinct populations: a brighter
population, with F814W $\sim 18-24$; and a fainter population, with
F814W $\gtrsim 24$. The brighter population is both significantly brighter and
significantly redder on average than the population of field sources. The
fainter population, on the other hand, is consistent in its magnitude and color
distribution with the majority of the faint blue field galaxy population.

Specifically, we have shown that this optically faint blue group is likely to
consist predominantly of low to moderate luminosity Type~2 AGN at reasonably
high redshifts for this class of object (at least $z \sim 1-2$). Their optical
properties would be dominated by their hosts, most probably late-type galaxies.
These Type~2 AGN are the only population of sources with a sufficiently high
ratio of X-ray to optical emission to be detectable in the 1~Msec catalog while
simultaneously displaying such faint optical magnitudes; other classes of
objects (including normal galaxies, starbursts, and unobscured AGN) are
observed at brighter optical magnitudes but would not be detectable in the
1~Msec catalog at these faint magnitude levels.

We have also discovered a faint, red X-ray source that is undetected in F606W
and has an F606W$\,-\,$F814W color at least 1~magnitude redder than the rest of
the faint X-ray emitting population. This object may also be a Type~2 AGN of
comparable luminosity to the others but located in a redder host galaxy, for
example an early-type galaxy or one that contains excessive amounts of dust.
However, this object may also be an example of a class of much more luminous
(hence more sparse) reddened, dusty QSOs located at higher redshift. Deep
spectroscopy should yield more clues about its nature.

In conclusion, we can state that the combination of deep Msec-level X-ray
surveys together with the unprecedented spatial resolution of {\sl HST},
are for the first time providing us with the ability to probe the faint end of
the obscured AGN luminosity function at cosmologically interesting redshifts.
This is supported by the relatively robust agreement between our observations
and models based on luminosity functions of AGN and normal galaxies. These
results show that we now have the capability to investigate in detail the
morphological properties of these sources and classify them accordingly. Our
{\sl HST} observations cover only $7.5\%$ of the full-depth 1~Msec CDFS area
(Giacconi et al. 2001b), thus there is clearly an enormous potential for
further observations to increase both the sample size and the depth at which
these objects are studied. Combined with deep spectroscopy on 8m-class
telescopes and studies at other wavebands (including, for example, mid-
to far-IR imaging with SIRTF), such samples will eventually be able to probe up
to and beyond the expected peak in AGN density evolution at $z \sim 2$, as well
as revealing their relationships to starbursts and other galaxy classes.

\acknowledgements
We gratefully acknowledge the award of {\sl HST} Director's Discretionary time
in support of this project. We also acknowledge support for this work which was
provided by NASA through GO grants GO-08809.01-A and GO-07267.01-A from the
Space Telescope Science Institute, which is operated by AURA, Inc., under NASA
Contract \hbox{NAS 5-26555}.
We thank the referee for very useful suggestions that helped to improve this
paper.

\clearpage

\begin{deluxetable}{cccccc}
\tablecaption{Observed Fields\label{tab:obs-summary}}
\tablewidth{0pt}
\tablehead{
\colhead{} & \colhead{R.A.} & \colhead{Dec.}
					& \colhead{F606W}
						& \colhead{F814W}
							& \colhead{+V3 Orientation} \\
\colhead{Field}
	& \colhead{(J2000)} & \colhead{(J2000)}
					& \colhead{Exp. Time (s)}
						& \colhead{Exp. Time (s)}
							& \colhead{(degrees)}
}
\startdata
CDFS1	& 03 32 28.80	& $-$27 45 53.0	& 3700	& 5800	& 65	\\
CDFS2	& 03 32 07.60	& $-$27 46 35.0	& 3700	& 5800	& 65	\\
CDFS3	& 03 32 12.70	& $-$27 43 02.0	& 3700	& 5800	& 65	\\
\enddata
\end{deluxetable}


\begin{deluxetable}{ccccc}
\tablecaption{Dither Strategy\label{tab:dither}}
\tablewidth{0pt}
\tablehead{
\colhead{Dither Offset}	& PC			& WF2		& WF3		& WF4
\\
\colhead{(arcseconds)}	& (pixels)		& (pixels)	& (pixels)	& (pixels)
}
\startdata
0\farcs00, 0\farcs00	& \phn\phn0, 0\phn\phn	& \phn\phn0, 0\phn\phn	& \phn\phn0, 0\phn\phn	& \phn\phn0, 0\phn\phn	\\
0\farcs50, 0\farcs25	& $-$11.0, $-$5.5\phn	& $-$2.5, 5.0\phn	& 5.0, 2.5		& \phn2.5, $-$5.0	\\
0\farcs75, 0\farcs75	& $-$16.5, $-$16.5	& $-$7.5, 7.5\phn	& 7.5, 7.5		& \phn7.5, $-$7.5	\\
0\farcs25, 0\farcs50	& \phn$-$5.5, $-$11.0	& $-$5.0, 2.5\phn	& 2.5, 5.0		& \phn5.0, $-$2.5
\enddata
\end{deluxetable}

\clearpage

\begin{deluxetable}{cccccc}
\tabletypesize{\scriptsize}
\tablecaption{{\sl HST}/WFPC2 Observation Log\label{tab:obs-detailed}}
\tablewidth{0pt}
\tablehead{
\colhead{Dither Offset}	& \colhead{WFPC2}
				& \colhead{Exp. Time}
					& \colhead{CDFS1}	&  \colhead{CDFS2}	& \colhead{CDFS3}
\\
\colhead{(arcseconds)}	& \colhead{Filter}
				& \colhead{(s)}
					& \colhead{Start Time (UT)}
								& \colhead{Start Time (UT)}
											& \colhead{Start Time (UT)}
}
\startdata
0\farcs00, 0\farcs00	& F814W	& 700	& 2000 Jul 22 19:14:15	& 2000 Jul 23 16:09:15	& 2000 Jul 27 19:56:15	\\
0\farcs00, 0\farcs00	& F814W	& 700	& 2000 Jul 22 19:31:15	& 2000 Jul 23 16:26:15	& 2000 Jul 27 20:13:15	\\
0\farcs00, 0\farcs00	& F606W	& 700	& 2000 Jul 22 19:48:15	& 2000 Jul 23 16:43:15	& 2000 Jul 27 20:30:15	\\
0\farcs00, 0\farcs00	& F606W	& 500	& 2000 Jul 22 20:48:15	& 2000 Jul 23 17:43:15	& 2000 Jul 27 21:28:15
\\[4pt]
0\farcs50, 0\farcs25	& F814W	& 800	& 2000 Jul 22 21:02:15	& 2000 Jul 23 17:57:15	& 2000 Jul 27 21:42:15	\\
0\farcs50, 0\farcs25	& F814W	& 800	& 2000 Jul 22 21:21:15	& 2000 Jul 23 18:16:15	& 2000 Jul 27 22:01:15	\\
0\farcs50, 0\farcs25	& F606W	& 400	& 2000 Jul 22 22:25:15	& 2000 Jul 23 19:20:15	& 2000 Jul 27 23:05:15	\\
0\farcs50, 0\farcs25	& F606W	& 400	& 2000 Jul 22 22:34:15	& 2000 Jul 23 19:29:15	& 2000 Jul 27 23:14:15
\\[4pt]
0\farcs75, 0\farcs75	& F814W	& 600	& 2000 Jul 22 22:46:15	& 2000 Jul 23 19:41:15	& 2000 Jul 27 23:26:15	\\
0\farcs75, 0\farcs75	& F814W	& 600	& 2000 Jul 22 22:59:15	& 2000 Jul 23 19:54:15	& 2000 Jul 27 23:39:15	\\
0\farcs75, 0\farcs75	& F606W	& 400	& 2000 Jul 23 00:05:15	& 2000 Jul 23 20:56:15	& 2000 Jul 28 00:43:15	\\
0\farcs75, 0\farcs75	& F606W	& 400	& 2000 Jul 23 00:14:15	& 2000 Jul 23 21:05:15	& 2000 Jul 28 00:52:15
\\[4pt]
0\farcs25, 0\farcs50	& F606W	& 400	& 2000 Jul 23 00:26:15	& 2000 Jul 23 21:17:15	& 2000 Jul 28 01:04:15	\\
0\farcs25, 0\farcs50	& F606W	& 500	& 2000 Jul 23 00:35:15	& 2000 Jul 23 21:26:15	& 2000 Jul 28 01:13:15	\\
0\farcs25, 0\farcs50	& F814W	& 800	& 2000 Jul 23 01:47:15	& 2000 Jul 23 22:33:15	& 2000 Jul 28 02:22:15	\\
0\farcs25, 0\farcs50	& F814W	& 800	& 2000 Jul 23 02:06:15	& 2000 Jul 23 22:52:15	& 2000 Jul 28 02:41:15
\enddata
\end{deluxetable}

\clearpage

\begin{deluxetable}{clcrrrrll}
\tabletypesize{\scriptsize}
\tablecaption{\label{tab:all_sources}%
Properties of the WFPC2 counterparts to the 1~Msec CDFS X-ray sources}
\tablewidth{0pt}
\tablehead{
\colhead{CDFS}
	& \colhead{IAU-Format}
			& \colhead{$(\Delta\alpha,\Delta\delta)$\tablenotemark{a}}
							& \colhead{F606W}
								& \colhead{F814W}
										& \colhead{$\log F_{\rm XS}$}
											& \colhead{$\log F_{\rm XH}$}
													& \colhead{$HR$}
														& \colhead{Optical} \\
\colhead{ID}
	& \colhead{Designation}
			& \colhead{(arcsec)}
							&  
								&  
									& \colhead{(cgs)}
											& \colhead{(cgs)}
													&	& \colhead{Morphology}
}
\startdata
\phn36	& J033233.1$-$274548 & $(+0.02,-0.03)$		& 22.61 & 21.63 & $-15.01$	& $-14.54$	& $-0.36$ & Spiral in CG \\
\phn38	& J033230.3$-$274505 & $(-0.01,-0.06)$		& 22.30 & 21.79 & $-14.31$	& $-14.05$	& $-0.56$ & Unresolved \\
\phn39	& J033230.1$-$274530 & $(+0.02,-0.02)$		& 21.45 & 21.36 & $-14.17$	& $-13.81$	& $-0.47$ & Unresolved \\
\phn52	& J033217.2$-$274304 & $(+0.15,-0.02)$		& 21.66 & 20.81 & $-14.41$	& $-14.13$	& $-0.54$ & Elliptical \\
\phn56	& J033213.3$-$274241 & $(+0.39,-0.18)$		& 20.85 & 20.00 & $-14.67$	& $-13.77$	& $+0.11$ & Ell., merger? \\
\phn58	& J033211.8$-$274629 & $(-0.05,-0.04)$		& 26.05 & 25.32 & $-15.21$	& $-14.73$	& $-0.34$ & LSB Spiral \\
\phn60	& J033211.0$-$274415 & $(+0.34,+0.03)$		& 22.50 & 21.96 & $-14.28$	& $-13.94$	& $-0.48$ & Unresolved \\
\phn61	& J033210.6$-$274309 & $(+0.27,+0.07)$		& 24.98 & 23.05 & $-14.35$	& $-13.96$	& $-0.44$ & Unresolved \\
\phn62	& J033209.5$-$274807 & $(-0.13,-0.03)$		& 20.98 & 20.55 & $-15.02$	& $-14.27$	& $-0.04$ & Unresolved \\
\phn63	& J033208.7$-$274735 & $(-0.05,+0.01)$		& 19.28 & 18.73 & $-13.42$	& $-13.10$	& $-0.49$ & Unresolved \\
\phn64	& J033208.1$-$274658 & $(-0.04,-0.04)$		& 25.39 & 25.17 & $-14.73$	& $-14.23$	& $-0.31$ & Spiral in group \\
\phn66	& J033203.7$-$274604 & $(+0.22,+0.08)$		& 21.45 & 20.16 & $-15.29$	& $-13.95$	& $+0.56$ & S0\\
\phn67	& J033202.5$-$274601 & $(+0.29,-0.06)$		& 24.81 & 23.62 & $-14.42$	& $-13.99$	& $-0.40$ & Elliptical \\
\phn78	& J033230.1$-$274524 & $(+0.01,+0.02)$		& 23.11 & 22.28 & $-14.78$	& $-14.50$	& $-0.54$ & Unresolved \\
\phn81	& J033226.0$-$274515 & $(-0.05,-0.03)$		& 26.56 & 26.24 & $-15.44$	& $-15.05$	& $-0.43$ & Indeterminate \\
\phn83	& J033215.0$-$274225 & $(+0.62,-0.63)$		& 23.64 & 22.65 & $-15.07$	& $-14.46$	& $-0.22$ & Elliptical \\
\phn86	& J033233.9$-$274521 & $(-0.15,-0.38)$		& 25.54 & 25.13 & $-15.93$	& $-15.16$	& $-0.04$ & LSB Irregular \\
\phn89	& J033208.3$-$274153 & $(+0.78,-0.44)$		& 25.32 & 25.20 & $-15.37$	& $-14.98$	& $-0.44$ & Indeterminate \\
   149	& J033212.3$-$274621 & $(-0.10,+0.15)$		& 23.86 & 22.55 & $-15.92$	& $-15.03$	& $+0.12$ & Spiral \\
   155	& J033208.0$-$274240 & $(+0.67,+0.06)$		& 22.80 & 21.73 & $-15.84$	& $-14.88$	& $+0.17$ & Spiral \\
   173	& J033216.8$-$274327 & $(-0.18,-0.35)$		& 22.97 & 22.01 & $-16.11$	& $<\!-15.5$	& $-1.00$ & Spiral \\
   185	& J033211.0$-$274343 & $(-0.29,-0.41)$		& 22.52 & 21.47 & $-16.12$	& $-15.19$	& $+0.14$ & Spiral \\
   224	& J033228.8$-$274621 & $(+0.11,+0.10)$		& 23.05 & 21.65 & $-15.62$	& $<\!-15.5$	& $-1.00$ & Elliptical \\
\phn226$\;$\tablenotemark{b}   %
	& J033204.5$-$274644 & \nodata			& $>$26.9 & $>$26.4 & $-15.34$	& $-14.93$	& $-0.40$ & \nodata \\
\phn227$\;$\tablenotemark{b}   %
	& J033205.4$-$274644 & \nodata			& $>$26.9 & $>$26.4 & $-16.00$	& $-14.66$	& $+0.57$ & \nodata \\
   266	& J033214.0$-$274249 & $(-0.15,-0.08)$		& 22.43 & 21.36 & $<\!-16.5$	& $-14.88$	& $+1.00$ & Elliptical \\
\phn515$\;$\tablenotemark{c}   %
	& J033232.2$-$274652 & $(+0.04,+0.02)$		& $>$27.9 & 26.41 & $-16.09$	& $-14.90$	& $+0.42$ & Indeterminate \\
\phn518$\;$\tablenotemark{b}   %
	& J033226.9$-$274605 & \nodata			& $>$26.9 & $>$26.4 & $-16.08$	& $-15.25$	& $+0.03$ & \nodata \\
   532	& J033214.2$-$274231 & $(-0.15,+0.28)$		& 24.89 & 24.51 & $-15.86$	& $-15.10$	& $-0.05$ & Indeterminate \\
   535	& J033211.5$-$274650 & $(-0.15,-0.34)$		& 22.81 & 21.74 & $-15.70$	& $-14.90$	& $+0.02$ & Elliptical \\
   536	& J033210.9$-$274235 & $(-0.29,+0.14)$		& 20.27 & 19.34 & $-15.76$	& $-15.17$	& $-0.24$ & Elliptical \\
   538	& J033208.6$-$274649 & $(+0.04,-0.01)$		& 19.82 & 18.94 & $-15.99$	& $-15.11$	& $+0.12$ & Spiral \\
   560	& J033206.3$-$274537 & $(+1.06,+0.37)$		& 22.85 & 21.42 & $-15.93$	& $<\!-15.5$	& $-1.00$ & Spiral \\
   563	& J033231.5$-$274624 & $(-0.05,+0.55)$		& 23.62 & 23.12 & $-16.22$	& $<\!-15.5$	& $-1.00$ & Spiral / Irr. \\
   593	& J033214.8$-$274403 & $(+1.32,+0.62)$		& 26.42 & 25.52 & $-16.08$	& $<\!-15.5$	& $-1.00$ & LSB Irregular \\
   594	& J033209.8$-$274249 & $(-0.33,+0.87)$		& 22.65 & 20.88 & $-15.48$	& $<\!-15.5$	& $-1.00$ & Elliptical \\
   623	& J033228.6$-$274659 & $(+0.73,-0.06)$		& 26.96 & 26.63 & $-16.11$	& $<\!-15.5$	& $-1.00$ & Indeterminate \\
   624	& J033229.3$-$274708 & $(-0.21,+0.05)$		& 22.61 & 21.06 & $-16.11$	& $<\!-15.5$	& $-1.00$ & Elliptical \\
   626	& J033209.5$-$274758 & $(+0.86,+0.84)$		& 25.71 & 25.26 & $-16.05$	& $<\!-15.5$	& $-1.00$ & LSB Irregular \\
   631	& J033215.2$-$274335 & $(-0.36,-0.08)$		& 25.26 & 23.85 & $-15.99$	& $<\!-15.5$	& $-1.00$ & Elliptical \\
\enddata
\tablenotetext{a}{J2000 Right Ascension $\alpha$ and Declination $\delta$
	offsets ({\sl HST}$-${\sl CXO}), in arcsec.}
\tablenotetext{b}{No counterpart within 2\arcsec\ of {\sl CXO} position;
	F606W, F814W mags are $3\sigma$ upper limits for the PC chip.}
\tablenotetext{c}{This object is detected in F814W on the WF4 chip with a
	magnitude F814W=26.41 (the limiting magnitude on the WF4 chip is 27.4);
	however it is not detected in F606W down to the $3\sigma$ F606W limiting
	magnitude of 27.9.}
\end{deluxetable}

\clearpage

\begin{figure}
\caption{\label{fig:dither}
Dither strategy employed in the {\sl HST}/WFPC2 observations of the three
fields in the CDFS. For each field, the observations were obtained using a
parallelogram dither pattern specified in the POS TARG reference frame of the
WFALL-FIX coordinate system (i.e., along the X,Y axes of the WF3 detector).
The measured shifts for each field location are shown in the top panels, along
with dashed lines indicating the commanded offsets. It can be seen that the
r.m.s. positioning accuracy of the large-scale offsets is generally well within
10~mas, while drifts at each location, from one exposure to the next, are
within 5~mas. The bottom panels display the shifts in terms of the fractional
sub-pixel offsets of the three WF chips (WF2, WF3, WF4). Circles correspond to
WF2, triangles to WF3 and squares to WF4. Open symbols are for the F606W
images, and filled symbols correspond to F814W. For example, a source initially
centered at the bottom-left quadrant of the pixel would move around to the
other three quadrants as shown. These figures show that the dither offsets have
generally been successful in achieving the desired 1/2-pixel subsampling along
both the X and Y axes of the detectors.}
\end{figure}


\begin{figure}
\caption{\label{fig:offsets}
(a)~Offsets in right ascension and declination to the nearest HST optical
counterpart, for each of the 40 X-ray sources in our three WFPC2 fields of
view. The three open squares represent sources for which the offset is
significantly larger than the 4-$\sigma$ statistical variation, and we thus
do not consider these as positive identifications. The other 37 X-ray sources
are all identified with optical counterparts, as discussed further in the text.
(b)~Offset as a function of off-axis angle on the Chandra ACIS-I chip. Our 3
WFPC2 fields cover a substantial range of off-axis angles; this plot shows that
the scatter of positional offsets remains approximately the same and does not
show a significant increase at larger off-axis angles.}
\end{figure}


\setcounter{figure}{2}
\begin{figure}
\caption{\label{fig:fullfields}
{\sl HST} greyscale images of each of the three WFPC2 fields in the CDFS, made
by combining the F606W and F814W datasets. North is to the top and East is to
the left. We also indicate the location of each of the X-ray sources from the
1~Msec Chandra catalog that are located within our WFPC2 field of view. The
X-ray source identifications correspond to the nomenclature of
	Giacconi et al. (2001b).
(a) This figure is for the CDFS1 field.}
\end{figure}


\setcounter{figure}{2}
\begin{figure}
\caption{(b) As for Fig.~\ref{fig:fullfields}a, but for the CDFS2 field.}
\end{figure}


\setcounter{figure}{2}
\begin{figure}
\caption{(c) As for Fig.~\ref{fig:fullfields}a, but for the CDFS3 field.}
\end{figure}



\setcounter{figure}{3}

\begin{figure}
\caption{\label{fig:smallfields_grey}
(a) {\sl HST} greyscale images of each of the X-ray sources in the three CDFS
WFPC2 fields, made by combining the F606W and F814W datasets. The X-ray
contours from the smoothed 1~Msec Chandra data are overlaid on each image (with
the first three contour levels corresponding to 3~sigma, and increasing
thereafter by a factor of two per contour level). For sources detected only
in one X-ray band (hard or soft), we show the contours from that band only;
these sources are indicated by upper limits in the corresponding X-ray band
in Table~\ref{tab:all_sources}. In all images, North is to the top and East is
to the left.}
\end{figure}


\setcounter{figure}{3}
\begin{figure}
\caption{(b) As for Fig.~\ref{fig:smallfields_grey}a.}
\end{figure}


\setcounter{figure}{3}
\begin{figure}
\caption{(c) As for Fig.~\ref{fig:smallfields_grey}a.}
\end{figure}


\setcounter{figure}{3}
\begin{figure}
\caption{(d) As for Fig.~\ref{fig:smallfields_grey}a.}
\end{figure}


\setcounter{figure}{3}
\begin{figure}
\caption{(e) As for Fig.~\ref{fig:smallfields_grey}a.}
\end{figure}



\begin{figure}
\caption{\label{fig:col_mag}
Color-magnitude diagram of the 3681 sources (small dots) with both
F606W and F814W detections in the three WFPC2 fields of this study.
Sources with X-ray emission detected in the 1~Msec CDFS image are
flagged with larger symbols according to their optical morphology:
resolved galaxies of types S0 and earlier (circles); resolved galaxies
of types Sa and later (squares); unresolved sources (stars); and
indeterminate (triangles). Their size indicates total X-ray flux, while their
shading represents X-ray hardness ratio (ranging from white to black for
$HR=-1$ to $HR=+1$ respectively).
We also show tracks for elliptical, spiral, starburst/irregular and Type~1
AGN template galaxy SEDs as a function of redshift; see text for further
details.
}
\end{figure}



\begin{figure}
\caption{\label{fig:smallfields_rgb}
{\sl HST} color images of each of the X-ray sources (made by assigning red to
F814W, blue to F606W, and green to the average of the two bands). Each image
is 8\arcsec\ on a side, with North to the top and East to the left. The objects
are generally arranged according ot their location on the color-magnitude plot
as shown in Figure~\ref{fig:col_mag}: bright sources near the top, and redder
sources toward the right. Some of the source catagories are also subdivided
according to their hardness ratio. The three non-detections are shown at the
bottom.}
\end{figure}



\begin{figure}
\caption{\label{fig:popns}
Histogram of the source counts at F814W, in bins of 0.5~magnitude, for the 37 optical
counterparts of the X-ray emitting sources, with curves overlaid showing the
contribution of various different classes of objects according to the models.
Note in particular that the spirals, ellipticals, and Type~1 AGN are all relatively
bright, while the Type~2 AGN are the only ones with a distribution that is peaked
fainter than F814W$\,\sim\,24$.}
\end{figure}


\clearpage

\begin{references}

\reference{} Abraham, R.~G., Ellis, R., Fabian, A., Tanvir, N., \&
        Glazebrook, K.~1999, MNRAS, 303, 641

\reference{} Alexander, ~D. ~M., Brandt, ~W. ~N., Hornschemeier, ~A. ~E.,
	Garmire, ~G. ~P., Schneider, ~D. ~P., \& Bauer, ~F. ~E. 2001, ApJ, in press
	(astroph/0107450)

\reference{} Bertin, E. \& Arnouts, S.~1996, A\&AS, 117, 393

\reference{} Biretta,~J.~A. et al.~2000, WFPC2 Instrument Handbook, Version 5.0
        (Baltimore: STScI)

\reference{} Brandt, ~W. ~N., Hornschemeier, ~A. ~E., Alexander, ~D. ~M.,
	Garmire, ~G. ~P., Schneider, ~D. ~P., Broos, ~P. ~S., Townsley, ~L. ~K.,
	Bautz, ~M. ~W., Feigelson, ~E. ~D., \& Griffiths, ~R. ~E. 2001, AJ 122, 1

\reference{} Boyle, B.~J., Shanks, T., Croom, S.~M., Smith, R.~J.,
        Miller, L., Loaring, N., \& Heymans, C.~2000, MNRAS, 317, 1014

\reference{} Casertano, S., de Mello, D., Dickinson, M., Ferguson, H. C.,
	Fruchter, A. S., Gonzalez-Lopezlira, R. A., Heyer, I., Hook, R. N.,
	Levay, Z., Lucas, R. A., Mack, J., Makidon, R. B., Mutchler, M.,
	Smith, T. E., Stiavelli, M., Wiggs, M. S., \& Williams, R. E.
	2000, AJ 120, 2747

\reference{} Coleman, G.~D., Wu., C.-C., \& Weedman, D.~W.~1980, ApJS, 43, 393

\reference{} Cristiani, S. \& Vio, R. 1990, A\&A 227, 385

\reference{} Della Ceca, R., Maccacaro, T., Rosati, P., \& Braito, V.
	2000, A\&A 355, 121

\reference{} Ferrarese, L. \& Merritt, D.~2000, ApJ, 539, 9

\reference{} Fruchter, A.~S. \& Hook, R.~2001, PASP, submitted

\reference{} Gebhardt, K., Kormendy, J., Ho, L. C., Bender, R., Bower,
        G., Dressler, A., Faber, S.~M., Filippenko, A.~V., Green, R.,
        Grillmair, C., Lauer, T. R., Magorrian, J., Pinkney, J., Richstone,
        D., \& Tremaine, S.~2000, ApJ, 539, 13

\reference{} Giacconi, R., Rosati, P., Tozzi, P., Nonino, M., Hasinger, G.,
        Norman, C., Bergeron, J., Borgani, S., Gilli, R., Gilmozzi, R., \&
        Zheng, W.~2001a, ApJ, in press (astro-ph/0007240)

\reference{} Giacconi, R., Zirm, A., Wang, J.~X., Rosati, P., Nonino, M.,
	Tozzi, P., Gilli, R., Mainieri, V., Hasinger, G., Kewley, L.,
        Bergeron, J., Borgani, S., Gilmozzi, R., Grogin, N., Koekemoer, A.,
        Schreier, E., Zheng, W. \& Norman, C.~2001b, ApJ, submitted

\reference{} Grogin, N.~A. et al. 2001, in preparation

\reference{} Hasinger, G., Burg, R., Giacconi, R., Schmidt, M., Truemper, J.,
        \& Zamorani, G.~1998, A\&A, 329, 482.

\reference{} Hasinger, G., et al.~1999, in "Highlights in X-ray Astronomy",
        MPE report 272, 199 (astro-ph/9901103)

\reference{} Hasinger, G., et al.~2001, in preparation

\reference{} Hornschemeier, A.~E., Brandt, W.~N., Garmire, G.~P.,
	Schneider, D.~P., Barger, A.~J., Broos, P.~S., Cowie, L.~L.,
	Townsley, L.~K., Bautz, M.~W., Burrows, D.~N., Chartas, G.,
	Feigelson, E.~D., Griffiths, R.~E., Lumb, D., Nousek, J.~A.,
	Ramsey, L.~W., Sargent, W.~L.~W.~2001, ApJ, 554, 742

\reference{} Koekemoer, A.~M., Gonzaga,~S., Fruchter,A., Biretta,~J.,
	Casertano,~S., Hsu,~J.-C., Lallo,~M., Mutchler,~M., \& Hook,~R.~2000,
	The HST Dither Handbook (Baltimore: STScI)

\reference{} Koekemoer, A.~M., et al.~2001, in preparation

\reference{} Lilly, S., Schade, D., Ellis, R., Le Fevre, O.,
        Brinchmann, J., Tresse, L., Abraham, R., Hammer, F., Crampton, D.,
        Colless, M., Glazebrook, K., Mallen-Ornelas, G., \& Broadhurst,
        T.~1998, ApJ, 500, 75

\reference{} Loveday, J., Peterson, B.~A., Efstathiou, G., \&
        Maddox, S.~J.~1992, ApJ, 390, 338

\reference{} Magorrian, J., Tremaine, S., Richstone, D., Bender, R.,
        Bower, G., Dressler, A., Faber, S.~M., Gebhardt, K., Green, R.,
        Grillmair, C., Kormendy, J., Lauer, T.~1998, AJ, 115, 2285

\reference{} Maiolino, R., Salvati, M., Antonelli, L., Comastri, A.,
        Fiore, F., Ghinassi, F., Gilli, R., La Franca, F., Mannucci, F.,
        Risaliti, G., Thompson, D., \& Vignali, C. 2000, A\&A, 355, L47

\reference{} Miyaji, T., Hasinger, G., \& Schmidt, M.~2000, A\&A, 353, 25

\reference{} Mushotzky, R., Cowie, L.~L., Barger, A.~J., \& Arnaud, K.~A.~2000,
        Nature, 404, 459

\reference{} Risaliti, G., Marconi, A., Maiolino, R., Salvati, M. \&
        Severgnini, P. 2001, A\&A, submitted (astro-ph/0102427)

\reference{} Rosati, P., Tozzi, R., Giacconi, R., Gilli, R., Hasinger, G.,
	Kewley, L., Mainieri, V., Nonino, M., Norman, C., Szokoly, G.,
	Wang, J.~X., Zirm, A., Bergeron, J., Borgani, S., Gilmozzi, R.,
	Grogin, N., Koekemoer, A., Schreier, E., \& Zheng, W. 2001, ApJ,
	submitted

\reference{} Saunders, W., Rowan-Robinson, M., Lawrence, A.,
        Efstathiou, G., Kaiser, N., Ellis, R.~S. \& Frenk, C.S.~1990, MNRAS,
        242, 318

\reference{} Schmidt, M., Giacconi, R., Hasinger, G., Truemper, J., \&
        Zamorani, G. 1998, in "Highlights in X-ray Astronomy", MPE report 272

\reference{} Schmitt, H.~R., Kinney, A.~L., Calzetti, D., \&
        Storchi-Bergmann, T.~1997, AJ 114, 592

\reference{} Schreier, E. J., Koekemoer, A. M., Grogin, N. A., Giacconi, R.,
	Gilli, R., Kewley, L.; Norman, C., Hasinger, G., Rosati, P.,
	Marconi, A., Salvati, M., \& Tozzi, P.
	2001, ApJ, submitted (astro-ph/0105271)

\reference{} Tozzi, P., Rosati, P., Nonino, M., Bergeron, J., Borgani, S.,
        Gilli, R., Gilmozzi, R., Hasinger, G., Grogin, N., Kewley, L.,
        Koekemoer, A., Norman, C., Schreier, E., Shaver, P., Szokoly, G.,
        Wang, J.~X., Zheng, W., Zirm, A. \& Giacconi, R.~2001, ApJ, submitted
        (astro-ph/0103014)

\reference{} Urry, C.~M. \& Padovani, P.~1995, PASP, 107, 803

\reference{} Voit et al. 1997, HST Data Handbook (Baltimore: STScI)

\reference{} Whitmore, B., Heyer, I., \& Casertano, S.~1999, PASP, 111, 1559

\reference{} Williams, R. E., Blacker, B., Dickinson, M., Dixon, W. V.,
	Ferguson, H. C., Fruchter, A. S., Giavalisco, M., Gilliland, R. L.,
	Heyer, I., Katsanis, R., Levay, Z., Lucas, R. A., McElroy, D. B.,
	Petro, L., Postman, M., Adorf, H.-M., \& Hook, R.
	1996, AJ 112, 1335

\end{references}
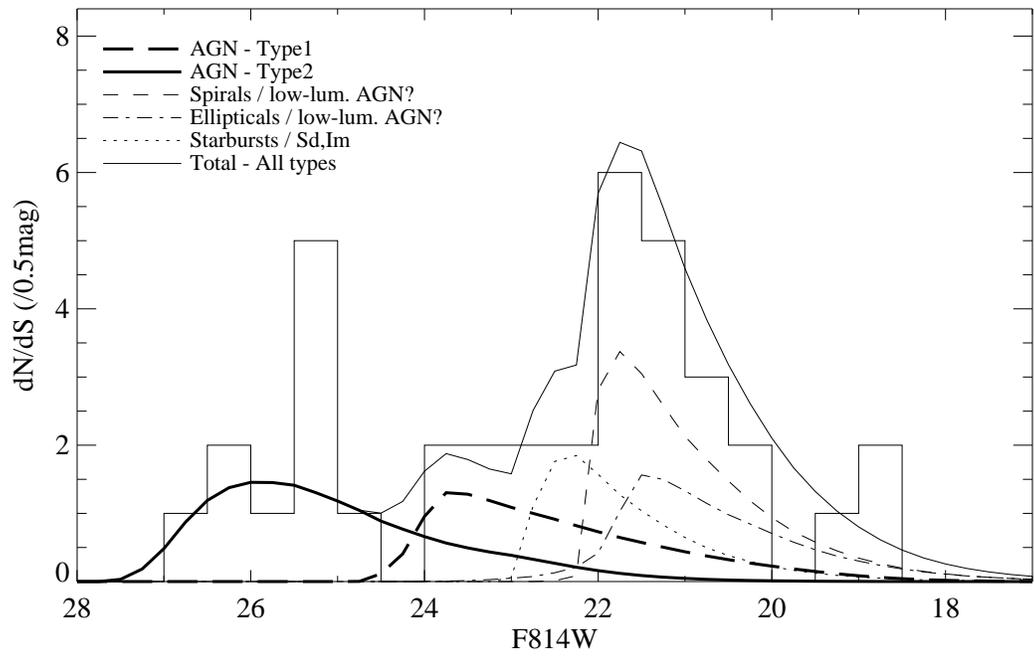
\end{document}